\documentclass{emulateapj}
\usepackage{lscape}
\newcommand{\msun}{\ensuremath{\rm M_\odot}}
\newcommand{\msunyr}{\ensuremath{\rm M_{\odot}\;{\rm yr}^{-1}}}
\newcommand{\Ha}{\ensuremath{\rm H\alpha}}

\newcommand{\Ntwo}{[\ion{N}{2}]}

\newcommand{\ztwo}{\ensuremath{z\sim2}}

\begin{document}

\title{\Ha\ OBSERVATIONS OF A LARGE SAMPLE OF GALAXIES AT
  \ztwo:\\  IMPLICATIONS FOR STAR FORMATION IN HIGH
  REDSHIFT GALAXIES\altaffilmark{1}}
\author{\sc Dawn
  K. Erb\altaffilmark{2}, Charles C. Steidel\altaffilmark{3}, Alice
  E. Shapley\altaffilmark{4},\\ Max Pettini\altaffilmark{5}, Naveen
  A. Reddy\altaffilmark{3}, Kurt L. Adelberger\altaffilmark{6}}

\shorttitle{STAR FORMATION IN $z\sim2$ GALAXIES}
\shortauthors{ERB ET AL.}
\slugcomment{Accepted for publication in \apj}

\altaffiltext{1}{Based on data obtained at the 
W.M. Keck Observatory, which is operated as a scientific partnership
among the California Institute of Technology, the University of
California, and NASA, and was made possible by the generous financial
support of the W.M. Keck Foundation.}  
\altaffiltext{2}{Harvard-Smithsonian Center for Astrophysics, MS 20,
  60 Garden St, Cambridge, MA 02138; derb@cfa.harvard.edu}
\altaffiltext{3}{California Institute of Technology, MS 105--24,
  Pasadena, CA 91125}  
\altaffiltext{4}{Department of Astrophysical Sciences, Princeton
  University, Peyton Hall, Ivy Lane, Princeton, NJ 08544}
\altaffiltext{5}{Institute of Astronomy, Madingley Road, Cambridge CB3
  0HA, UK}
\altaffiltext{6}{McKinsey and Company, 1420 Fifth Avenue, Suite 3100,
  Seattle, WA, 98101}

\begin{abstract}
Using \Ha\ spectra of 114 rest-frame UV-selected galaxies at \ztwo, we
compare inferred star formation rates (SFRs) with those determined
from the UV continuum luminosity.  After correcting for extinction
using standard techniques based on the UV continuum slope, we find
excellent agreement between the indicators, with $\langle \rm
SFR_{\Ha} \rangle = 31$ \msunyr\ and $\langle \rm SFR_{UV} \rangle =
29$ \msunyr.  The agreement between the indicators suggests that the
UV luminosity is attenuated by an typical factor of $\sim4.5$ (with a
range from no attenuation to a factor of $\sim100$ for the most
obscured object in the sample), in good agreement with estimates of
obscuration from X-ray, radio and mid-IR data.  The \Ha\ luminosity is
attenuated by a factor of $\sim1.7$ on average, and the maximum
\Ha\ attenuation is a factor of $\sim5$.  In agreement with X-ray and
mid-IR studies, we find that the SFR increases with increasing stellar
mass and at brighter $K$ magnitudes, to $\langle \rm SFR_{\Ha} \rangle
\sim 60$ \msunyr\ for galaxies with $K_s<20$; the correlation between
$K$ magnitude and SFR is much stronger than the correlation between
stellar mass and SFR.  All galaxies in the
sample have SFRs per unit area $\Sigma_{\rm SFR}$ in the range
observed in local starbursts.  We compare the instantaneous SFRs and
the past average SFRs as inferred from the ages and stellar masses,
finding that for most of the sample, the current SFR is an adequate
representation of the past average.  There is some evidence that the
most massive galaxies ($M_{\star}>10^{11}$ \msun) have had higher SFRs
in the past.
\end{abstract}


\section{Introduction}
Recent studies have indicated that a large fraction of the stellar
mass in the universe today formed at $z>1$ \citep{dpfb03,rrf+03}.
Thus it is especially important to understand the rates and timescales
of star formation in galaxies at high redshift.  Effective techniques
now exist for the selection of galaxies at \ztwo; these use the
galaxies' observed optical \citep{ssp+04} or near-IR (DRGs, or Distant
Red Galaxies with $J-K>2.3$; \citealt{flr+03}) colors, or a
combination of the two ($BzK$-selected galaxies, \citealt{bzk}), and
can be used to select both star-forming and passively evolving
galaxies.  Galaxies selected by their optical ($U_nG{\cal R}$) colors
comprise $\sim70$\% of the star formation rate density at \ztwo\
(including $U_nG{\cal R}$ and $BzK$ galaxies to $K=22$ and DRGs to
$K=21$), and range in bolometric luminosity from $\sim10^{10}$
L$_{\odot}$ to $>10^{12}$ L$_{\odot}$ \citep{res+05,rsf+06}.  This
paper focuses on the star formation properties of such galaxies.

Advances in instrumentation have enabled the determination of star
formation rates at an increasing range of wavelengths.  The most
straightforward data to obtain are optical images, which sample the
rest-frame UV at \ztwo.  The UV light is attenuated by dust, however,
and the magnitude of this extinction must be understood in order to
obtain accurate SFRs.  At high redshift, extinction is most readily
determined by the UV slope in combination with an extinction law such
as that of \citet{cab+00}; such an approach has been found to
represent adequately the average extinction of most \ztwo\ star-forming
galaxies, although the UV slope may overpredict the extinction for the
youngest objects and underpredict it for the reddest and dustiest
galaxies \citep{rs04,rsf+06,pmd+05}.

The UV light which is absorbed by dust is reradiated in the infrared,
and thus the FIR luminosity provides a more direct estimate of the
bolometric star formation rate for many galaxies.  The FIR light can be directly
detected at submillimeter wavelengths for only the most luminous
\ztwo\ galaxies (e.g.\ \citealt{cbsi05}), but it is possible to make
use of correlations between the FIR and X-ray and radio emission to
estimate SFRs for more typical galaxies \citep{rs04}.  Such average
star-forming galaxies at \ztwo\ are not detected even in the deepest
X-ray and radio images, however, so these techniques work primarily for
stacked images which give only the average SFR of a sample.  More
recently, the Spitzer Space Telescope has enabled the detection of the
rest-frame IR light from individual \ztwo\ galaxies, for the most
direct determinations of bolometric SFRs \citep{pmd+05,rsf+06}.  Such
estimates are still somewhat indirect, requiring
templates to convert from the observed 5--8.5 \micron\ luminosity to
the total infrared luminosity $L_{\rm IR}$; however, these conversions give
good average agreement with X-ray and dust-corrected UV estimates of
SFRs. 

One of the most widely used star formation indicators in local
galaxies is the \Ha\ emission line, which traces the formation of
massive stars through recombination in \ion{H}{2} regions.  This is
one of the most instantaneous measures of the SFR, and it has the
advantage of being particularly well-calibrated
(e.g.\ \citealt{k98,bcw+04}).  However, it is much more difficult to
apply at high redshift because the \Ha\ line shifts into the near-IR
for $z\gtrsim0.5$.  Previous studies of \Ha-determined SFRs at high
redshift have therefore been limited to relatively small samples of a
few to $\sim20$ galaxies \citep{ess+03,vff+04,ssc+04}.  Regardless of
sample size, these studies have demonstrated that the detection of
\Ha\ at \ztwo\ is quite feasible with an 8--10 m telescope for
galaxies with SFRs greater than a few \msunyr.

A galaxy's star formation history is as important as its current star
formation rate, but is considerably more difficult to determine.  The
time at which galaxies begin forming stars is fundamental to models of
galaxy formation, and so we would like to know the ages of galaxies,
both locally and at high redshift, and whether or not their current
SFRs are representative of their past star formation rates.
Constraints on the histories of galaxies can be obtained by modeling
their integrated light as the sum of stellar populations of varying
ages.  This has been done for large samples of local galaxies
(e.g.\ \citealt{bcw+04,hpjd04}), but at high redshifts it is found
that population synthesis models with a variety of simple star
formation histories provide adequate fits to the broadband SEDs
\citep{pdf01,ssa+01,sse+05}.  With a sufficiently large sample,
however, statistically meaningful results may still be obtained.  Here
we take advantage of the largest sample of \Ha\ spectra yet assembled
at high redshift, in combination with stellar masses and ages from
population synthesis modeling, to compare the current star formation
rate with the estimated past average.

This paper is one of several presenting the analysis of the
\Ha\ spectra of 114 \ztwo\ galaxies selected by their rest-frame UV
colors.  The paper is organized as follows.  In \S\ref{sec:sample} we
describe the selection of our sample, the observations, and our data
reduction procedures.  We briefly outline the modeling procedure by
which we determine stellar masses and other stellar population
parameters in \S\ref{sec:smass}.  In \S\ref{sec:sf} we calculate and
compare star formation rates from \Ha\ and rest-frame UV luminosities.
\S\ref{sec:sftime} discusses constraints on timescales for star
formation.  We summarize our results in \S\ref{sec:conclude}.
Separately, \citet{ess+06} focus on the galaxies' kinematics and on
comparisons of stellar, dynamical and inferred gas masses, and
\citet{esp+06} use the same sample of \Ha\ spectra to construct composite
spectra according to stellar mass to show that there is a strong
correlation between increasing oxygen abundance as measured by the
\Ntwo/\Ha\ ratio and increasing stellar mass.  Galactic
outflows in this sample are discussed by Steidel et al.\ (2006, in
preparation).

A cosmology with $H_0=70\;{\rm km}\;{\rm s}^{-1}\;{\rm Mpc}^{-1}$,
$\Omega_m=0.3$, and $\Omega_{\Lambda}=0.7$ is assumed throughout.  In
such a cosmology, 1\arcsec\ at $z=2.24$ (the mean redshift of the
current sample) corresponds to 8.2 kpc, and at this redshift the
universe is 2.9 Gyr old, or 21\% of its present age.  
 
\section{Sample Selection, Observations and Data Reduction}
\label{sec:sample}

The selection of the sample and our observing and data reduction
procedures are described in detail by \citet{ess+06}.  We summarize
the object selection briefly here.  The galaxies discussed in this
paper are drawn from the rest-frame UV-selected $z\sim2$ spectroscopic
sample described by \citet{ssp+04}.  The candidate galaxies are
selected by their $U_nG\cal R$ colors (from deep optical images
discussed by \citealt{ssp+04}), with redshifts then confirmed in the
rest-frame UV using the LRIS-B spectrograph on the Keck I telescope.
Galaxies were selected for \Ha\ observations for a wide variety of
reasons, and the \Ha\ sample is not necessarily representative of the
UV-selected sample as a whole; because we selected some galaxies based
on their bright $K$ magnitudes or red ${\cal R}-K$ colors, and because
our \Ha\ detection rate is lower for galaxies that are very faint in
$K$ (as discusssed in more detail by \citealt{ess+06}), the
\Ha\ sample is slightly more massive on average than the UV-selected
\ztwo\ sample as a whole, though it spans the full range of properties
covered by the total sample. The galaxies observed are listed in Table 1;
their coordinates and photometric properties are given in Table 1 of
\citet{ess+06}.

For the purposes of comparisons with other surveys, 10 of the 87
galaxies for which we have \Ha\ spectra and $JK_s$ photometry have
$J-K_s>2.3$ (the selection criterion for the FIRES survey,
\citealt{flr+03}); this is similar to the $\sim$12\% of UV-selected
galaxies which meet this criterion \citep{res+05}.  18 of the 93
galaxies for which we have $K$ magnitudes have $K_s<20$, the selection
criterion for the K20 survey \citep{cdm+02}; this is a higher fraction
than is found in the full UV-selected sample ($\sim10$\%), because we
intentionally targeted many $K$-bright galaxies \citep{sep+04}.  Five
of the 10 galaxies with $J-K_s>2.3$ also have $K_s<20$.

\subsection{Near-IR Spectra}
\label{sec:irspec}
The \Ha\ spectra were obtained with the near-IR spectrograph NIRSPEC
\citep{mbb+98} on the Keck II telescope in low-resolution
($R\sim1400$) mode, and reduced using the standard procedures
described by \citet{ess+03}.  We comment here on the flux calibration,
which is the most difficult step in the process but is essential to
the determination of star formation rates.  Absolute flux calibration
is subject to significant uncertainties, primarily due to slit losses
from the seeing and imperfect centering of the object on the slit
(objects are acquired via blind offsets from a nearby bright star).
Because the exposures of the standard stars used as reference are not
usually taken immediately before or after the science targets
(primarily because the NIRSPEC detector suffers from charge
persistence after observations of bright objects), the calibration may
also be affected by differences in seeing and weather conditions
between the science and calibration observations.

Several methods have been used to assess the accuracy of
the flux calibration.  Using a narrow-band image of the Q1700 field
(centered on \Ha\ at $z=2.3$, for observations of the proto-cluster
described by \citealt{sas+05}), we have measured narrowband
\Ha\ fluxes for six of the objects in our sample, and find that the
NIRSPEC \Ha\ fluxes are $\sim50$\% lower.  For those (relatively few)
galaxies for which we detect significant continuum in the NIRSPEC
spectra, we can compare the average flux density in the $K$ band with
the broadband magnitudes.  These tests indicate that the NIRSPEC
fluxes are low by a factor of $\sim2$ or more.  We have also assessed
the effects of losses from the slit and the aperture used to extract
the spectra by constructing a composite two-dimensional spectrum of
all the objects in the sample and comparing its spatial profile to the
widths of the slit and our aperture.  This test indicates losses of
$\sim40$\%, although this figure represents a lower limit because our
procedure of dithering the object along the slit and subtracting
adjacent images results in the occasional loss of flux from extended
wings.  

Motivated by these tests, we have when noted applied a factor of two
aperture correction for the determination of star formation rates and
\Ha\ equivalent widths.  The correction is imprecise, as the fraction
of flux lost undoubtedly varies from object to object, but application
of the correction results in a closer approximation to the true
average flux of the sample than leaving the fluxes uncorrected (as
shown by the good agreement obtained between \Ha\ SFRs and those
determined at other wavelengths).

\subsection{Near-IR and Mid-IR Imaging}
We also make use of $J$-band and $K_s$-band images obtained with the
Wide-field IR Camera (WIRC; \citealt{wirc}) on the 5-m Palomar Hale
telescope, and mid-IR images from the Infrared Array Camera (IRAC) on
the Spitzer Space Telescope.  These data and our reduction procedures
are described by \citet{ess+06}.

\section{Model SEDs and Stellar Masses}
\label{sec:smass}

We determine best-fit model SEDs and stellar population parameters for
the 93 galaxies for which we have $K$-band magnitudes.  Most of these
(87) also have $J$-band magnitudes, and 35 (in the GOODS-N and Q1700
fields) have been observed at rest-frame near-IR wavelengths with
IRAC.  We use a modeling procedure identical to that described in
detail by \citet{sse+05}, with the exception that we employ a
\citet{c03} initial mass function (IMF) rather than the \citet{s55}
IMF used by \citet{sse+05}.  This results in stellar masses and star
formation rates 1.8 times lower.

The method is reviewed by \citet{ess+06}, and the results are
presented in Table 2 of that paper.  Using the solar metallicity
\citet{bc03} population synthesis models and a variety of simple star
formation histories of the form ${\rm SFR} \propto e^{(-t_{\rm
    sf}/\tau)}$, with $\tau=10$, 20, 50, 100, 200, 500, 1000, 2000 and
5000 Myr, as well as $\tau=\infty$ (i.e.\ constant star formation,
CSF), we determine the values of the age, $E(B-V)$ (using the
\citealt{cab+00} extinction law), star formation rate and stellar mass
which best match the observed 0.3--8 \micron\ photometry.  The mean
stellar mass is $3.6 \times 10^{10}$ \msun, and the median is $1.9
\times 10^{10}$ \msun.  The mean age is 1046 Myr, and the median age
is 570 Myr.  The sample has a mean $E(B-V)$ of 0.16 and a median of
0.15.  The mean SFR is 52 \msun\ yr$^{-1}$, while the median is 23
\msunyr; the difference between the two reflects the fact that a few
objects are best fit with high SFRs ($>300$ \msunyr).  We determine
uncertainties through a series of Monte Carlo simulations which
account for photometric uncertainties and degeneracies between age,
reddening and star formation history.  The simulations are described
by \citet{sse+05}.  The resulting mean fractional uncertainties are
$\langle \sigma_x/\langle x\rangle \rangle= 0.7$, 0.5, 0.6 and 0.4 in
$E(B-V)$, age, SFR and stellar mass respectively.  We also briefly
consider two-component models, to assess the possible presence of an
older stellar population hidden by current star formation.  As
discussed in more detail by \citet{ess+06}, we find that the data do
not favor large amounts of hidden mass; the most plausible of the
two-component models increase the total stellar masses by a factor of
$\sim2$--3, comparable to the uncertainties in the single component
modeling.

\section{Star Formation Rates}
\label{sec:sf}
There are three methods of estimating star formation rates for most of
the galaxies in the sample.  In addition to the \Ha\ luminosity, which
will be used for the primary analysis in this paper, SFRs can be
calculated from the rest-frame UV continuum and the normalization of
the best-fit model SED (see \S\ref{sec:smass}; these last two methods
both use the UV continuum, so are not independent).  The
correspondence of \Ha\ luminosity with SFR in particular is especially
useful because it is widely used in the local universe and has
recently been studied in detail using large samples of galaxies from
the SDSS \citep{hmn+03,bcw+04}.  \Ha\ also provides a nearly
instantaneous meaure of the SFR, because only stars with masses
greater that 10 \msun\ and ages less than 20 Myr contribute
significantly to the ionizing flux.  We use the \citet{k98}
transformation between \Ha\ luminosity and SFR, which assumes case B
recombination, a Salpeter IMF ranging from 0.1 to 100 \msun\ which we
convert to a Chabrier IMF by dividing the SFRs by 1.8, and that all
the ionizing photons are reprocessed into nebular line emission.
Using maximum likelihood SFRs from the full set of nebular emission
lines, \citet{bcw+04} show that this approximation works well for an
average star-forming galaxy, but that massive, metal rich galaxies
produce less \Ha\ luminosity for the same SFR than low mass, metal
poor galaxies.  This is probably a metallicity effect, as increased
line blanketing in metal-rich stars decreases the number of ionizing
photons.  The galaxies studied here follow a trend similar to local
galaxies in mass and metallicity, though probably offset to lower
metallicities at a given stellar mass \citep{esp+06}.  The largest
dispersion in the conversion factor from \Ha\ luminosity to star
formation rate is found for the most massive and metal-rich local
galaxies (see Figure 7, \citealt{bcw+04}); if our sample does not
contain galaxies with the highest metallicities observed in the local
universe, then the dispersion in the conversion factor is probably
less than our uncertainties from other sources, though we may be
biased toward overestimating the SFR by $\sim0.1$ dex.

In order to calculate SFRs from the UV continuum we use the observed
$G$-band magnitude, which corresponds to a mean rest-frame wavelength
of 1480 \AA\ for the galaxies in our sample (except for the 5 galaxies
at $z\sim1.5$, for which the $U_n$ magnitude corresponds to $\sim1500$
\AA).  We use the \citet{k98} conversion between 1500 \AA\ luminosity
and SFR, which assumes a timescale of $\sim10^8$ years for the galaxy
to reach its full UV continuum luminosity.  Because \Ha\ is sensitive
to only the most massive stars, it is a more instantaneous measure of
SFR than the UV continuum; however, for a constant SFR the continuum
luminosity rises by a factor of only 1.6 between ages of 10 and 100
Myr, so even for the youngest objects the UV continuum will not
severely underestimate the SFR.  We again convert from a Salpeter to
a Chabrier IMF.

\begin{figure*}[htbp]
\plotone{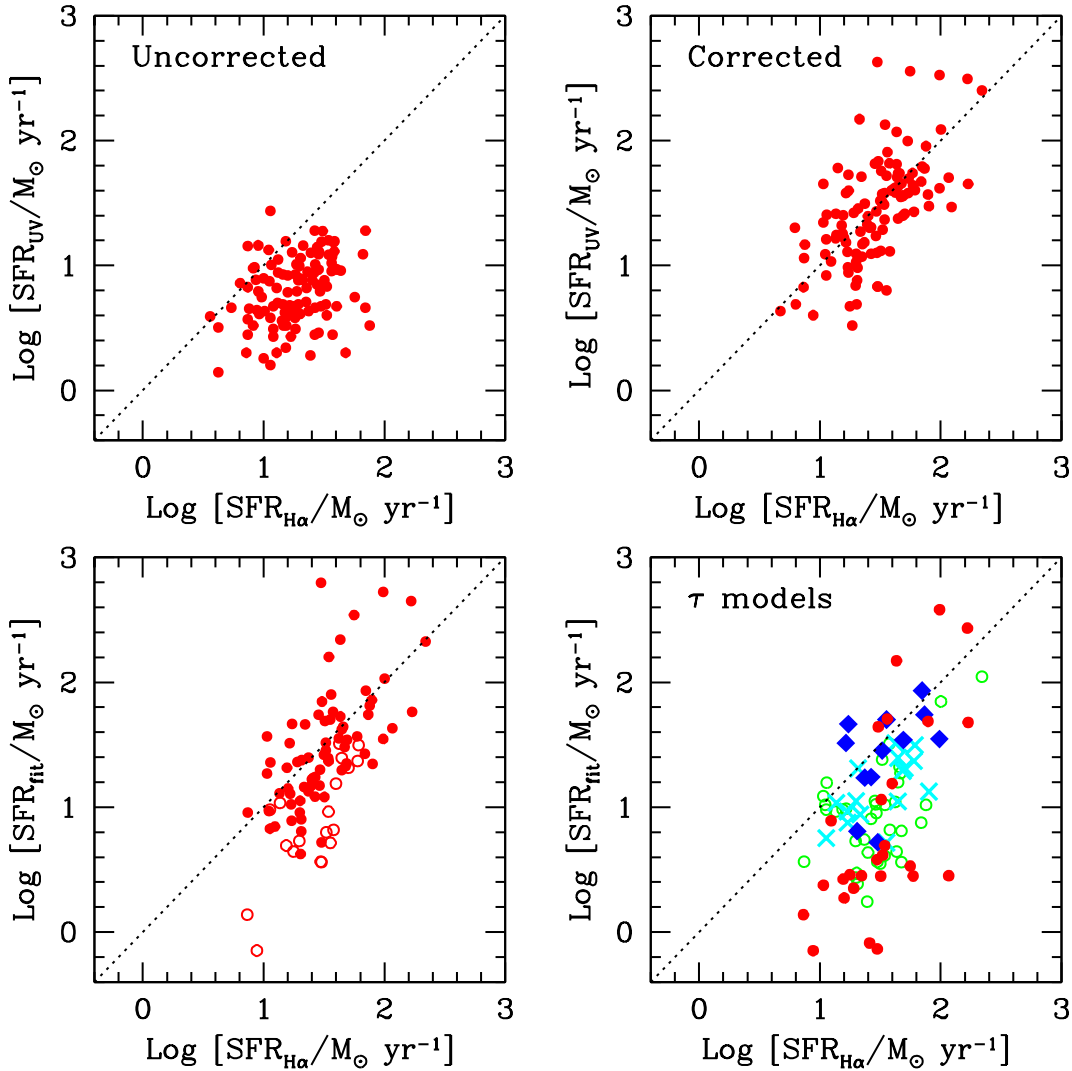}
\caption{A comparison of star formation rates from \Ha, the UV
  continuum, and the SED fits.  Upper left: SFR$_{\Ha}$ vs.\ SFR$_{\rm
    UV}$, without correcting for extinction.  Upper right: SFR$_{\Ha}$
  vs.\ SFR$_{\rm UV}$, with both SFRs corrected for extinction.  Lower
  left: Corrected SFR$_{\Ha}$ vs.\ the SFR obtained from the
  normalization of the adopted model SED.  Solid symbols are constant
  star formation models, and the open symbols represent objects for
  which we have adopted a model with an exponentially decreasing star
  formation rate.  Lower right: Corrected SFR$_{\Ha}$ vs.\ the SFR of
  the best-fitting declining model for each object.  Filled red circles are
  galaxies with $\tau$=10, 20 or 50 Myr, open green circles have
  $\tau$=100, 200 or 500 Myr, cyan crosses have $\tau$=1, 2 or 5 Gyr,
  and blue diamonds are constant star formation models.  The use of
  steeply declining $\tau$ models decreases the SFR with respect to
  that found from \Ha.}
\label{fig:sfrcomp}
\end{figure*}

We compare the various SFRs in Figure~\ref{fig:sfrcomp}.  The upper
left panel shows $\rm SFR_{UV}$ vs. $\rm SFR_{\Ha}$, without
correcting for extinction (in all cases we apply a factor of two
aperture correction to the \Ha\ SFRs, as discussed in
\S\ref{sec:irspec}).  There is considerable scatter, but the
probability that the data are uncorrelated is $P=0.0006$, for a
significance of the correlation of 3.4$\sigma$.  We find a mean and
standard deviation $\langle \rm SFR_{\Ha} \rangle = 22 \pm 14$ \msunyr,
and $\langle \rm SFR_{UV} \rangle = 8 \pm 5$ \msunyr.  In the upper
right panel both fluxes have been corrected for extinction, using the
\citet{cab+00} extinction law and the best-fit values of $E(B-V)$ from
the SED fits.  For those galaxies which do not have SED fits because
we lack the $K$ magnitude, $E(B-V)$ is calculated from the UV
continuum slope as measured by the $G-{\cal R}$ color, assuming a 570
Myr old SED with constant star formation; this is the median best-fit
age of the current sample.  The value of $E(B-V)$
calculated from the  $G-{\cal R}$ color in this way changes
by less than 10\% for assumed ages from 300--1000 Myr, though for
young objects $E(B-V)$ will probably be underestimated using this
method.  The value of $E(B-V)$ used for each galaxy is shown in
Table~\ref{tab:sfr}; the mean value is $\langle E(B-V) \rangle =
0.16$.  We have used the same value of $E(B-V)$ for the stellar UV
continuum and for the nebular emission lines, rather than $E(B-V)_{\rm
  stellar}=0.4 E(B-V)_{\rm neb}$ as proposed by \citet{cab+00},
because the latter assumption significantly overpredicts the \Ha\ SFRs
with respect to the UV SFRs.  The relative extinction suffered by the
stellar continuum and the nebular emission lines is an additional
source of uncertainty in our SFRs.  After the above corrections, we
find a mean and standard deviation $\langle \rm SFR_{\Ha} \rangle = 31
\pm 18$ \msunyr, and $\langle \rm SFR_{UV} \rangle = 29 \pm 19$
\msunyr, using 3$\sigma$ rejection to compute the statistics in order
to prevent the few objects with very high SFRs (particularly from the UV
luminosity) from biasing the distribution.  

The correlation between the corrected \Ha\ and UV SFRs is highly
significant (6.8$\sigma$), with an rms scatter of 0.3 dex.  Some of
this correlation may be due to the extinction correction applied to
both SFRs; to test the significance of this effect, we have randomized
the lists of uncorrected \Ha\ and UV fluxes to create many sets of
mismatched pairs, and applied the same (also randomized) value of
$E(B-V)$ to both fluxes in each pair.  In 10,000 trials we never
observe a correlation as strong as that observed in the real data; the
average trial has a correlation significance of 2.8$\sigma$ induced by
the extinction correction.  The much higher correlation significance in
the real data confirms the underlying correlation of the uncorrected SFRs.

In the lower panels of Figure~\ref{fig:sfrcomp} we compare the
corrected \Ha\ SFRs with those determined by the normalization of the
best-fitting SED.  The SED modeling uses the extinction-corrected UV
luminosity to determine SFRs, as we have done more directly in the
comparison discussed above; the difference is that the modeling
includes a variety of star formation histories.  The primary
purpose of this comparison is therefore to assess the effect of the
assumed star formation history on SFRs determined from SED modeling.
The lower left panel shows the SFR of our adopted best-fit model
vs. $\rm SFR_{\Ha}$.  The correlation is strong (5.3$\sigma$)
and the rms scatter is 0.3 dex.  The mean SFR from the SED fits is
$\langle \rm SFR_{\rm fit} \rangle = 24 \pm 17$ \msunyr, again
computed with 3$\sigma$ rejection because of the few objects with very
high SFRs. 70\% of the objects have $\rm SFR_{\Ha}>SFR_{fit}$.  The
points with open circles are those for which we have used models with
exponentially declining SFRs (SFR $\propto e^{-t/\tau}$) because they
provided a significantly better fit than the constant star formation
models; it is clear that the use of declining models depresses the
SFR.  This can be seen further in the lower right panel of
Figure~\ref{fig:sfrcomp}, in which we plot the SFR of the best-fitting
declining model vs. $\rm SFR_{\Ha}$.  The points are coded according
to the value of $\tau$: filled red circles are those galaxies best fit
with $\tau$=10, 20 or 50 Myr models, open green circles have
$\tau$=100, 200 or 500 Myr, cyan crosses have $\tau$=1, 2 or 5 Gyr,
and blue diamonds are constant star formation models.  As expected,
the steeply declining $\tau$ models yield the lowest SFRs, since they
allow the SFR to drop significantly during the lifetime of massive
stars. The objects with the highest SFRs are also formally best fit by
steeply declining models; these are generally young, highly reddened
objects that are acceptably fit by all values of $\tau$ and have high
SFRs for all star formation histories.  It is important to bear in
mind when considering the $\tau$ models that they are undoubtedly an
oversimplification of the likely star formation histories.  A model
with declining star formation may be required to obtain an acceptable
fit when a galaxy shows significant light from a previous generation
of stars as well as a current star formation episode, even if the
current episode is best described by constant star formation.  In such
cases the current SFR is likely to be underestimated.  Two component
models which decouple the current star formation episode from the
older population are more successful in determining current SFRs;
general two-component models that add a linear combination of a
current episode of constant star formation and an old burst (as
described by \citealt{ess+06}) are significantly better at matching
the \Ha-determined SFRs of galaxies which require $\tau$ models, while
still providing an acceptable fit to the SED.

We conclude that a typical galaxy in our sample has a star formation
rate of $\sim30$ \msunyr, though the SFRs of individual objects vary
from $\sim7$ to $\sim200$ \msunyr.  The dispersion in the correlations
suggest an uncertainty of a factor of $\sim2$ for individual galaxies,
as expected given the uncertainty of the aperture correction on
individual objects.  This result is in very good agreement with the
mean SFR of $\sim28$ \msunyr\ determined for the \ztwo\ UV-selected
sample from X-ray stacking techniques (\citealt{rs04,res+05}; we have
converted their value to a Chabrier IMF for comparison with our
sample).  We also find good agreeement between the \Ha\ SFRs and those
determined from 24 \micron\ observations; \citet{rsf+06} show that for
$\sim10$ galaxies in the GOODS-N field, the bolometric luminosities
implied by the corrected \Ha\ SFRs agree well with those inferred from
the 24 \micron\ luminosity.

\begin{figure}[htbp]
\plotone{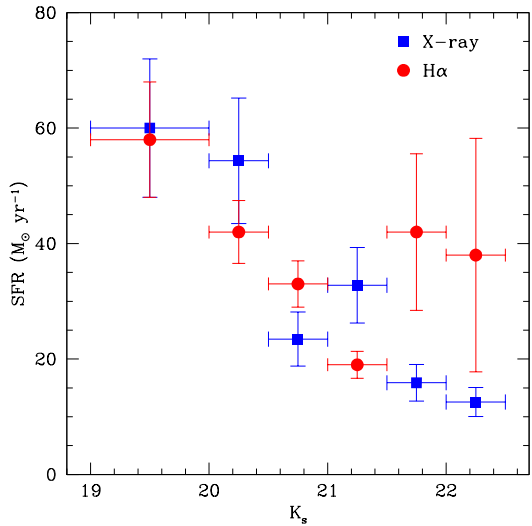}
\caption{Star formation rates from \Ha\ and X-ray stacking, as a
  function of $K$ magnitude.  Red circles from left to right represent
  the average extinction-corrected SFR$_{\Ha}$ of galaxies with
  $19<K_s\leq20$ and in 0.5 magnitude bins between $K_s=20$ and
  $K_s=22.5$, excluding AGN.  The average SFRs determined by stacking
  deep X-ray images of a slightly overlapping sample of
  \ztwo\ galaxies in the GOODS-N field in the same ranges of $K$
  magnitude are shown by the blue squares \citep{res+05}.  The upturn
  in SFR$_{\Ha}$ at faint $K$ magnitudes is probably a selection
  effect, because we are less likely to detect \Ha\ in galaxies faint
  in $K$ and because X-rays may underestimate the SFRs for young
  objects.}
\label{fig:haxray}
\end{figure}

A further result of the \citet{res+05} study is that the SFR increases
with increasing $K$-band luminosity.  We compare the current sample to
the results of \citet{res+05} by dividing our sample (excluding AGN)
into bins in $K$ magnitude and finding the average corrected
SFR$_{\Ha}$ in each bin.  The results are shown in
Figure~\ref{fig:haxray}, where the red circles are the average
\Ha\ SFRs and the blue squares are the SFRs from the X-ray stacking of
\citet{res+05}.  This is a comparison of similar objects, but not the
same objects; the X-ray data are available only in the GOODS-N field,
so the overlap between the two samples is small.  The agreement is
quite good for objects with $K\lesssim21$, but the \Ha\ data shows a
rise in SFR for $K$-faint, low stellar mass objects that is not seen
in the X-ray sample.  This discrepancy is likely related to at least
two different selection effects which complicate the comparison of
SFRs at faint $K$ magnitudes.  As noted in \S\ref{sec:sample} and
discussed in more detail by \citet{ess+06}, we are less likely to
detect \Ha\ emission for objects that are faint in $K$, unless they
have high SFRs.  Factoring in non-detections of $K$-faint galaxies
would probably lower the two right-most points considerably (we have
not done this because of the difficulty in distinguishing
non-detections due to low flux levels from non-detections for other
reasons).  If the low stellar mass objects in the \Ha\ sample are
young starbursts, the relative timescales of X-rays and \Ha\ as star
formation rate indicators may also be a factor.  The \Ha\ luminosity
is nearly instantaneous, while the X-ray luminosity increases for the
first $\sim10^8$ years as O/B stars die and become high-mass X-ray
binaries.  The X-rays may thus underestimate the SFR for very young
objects.  Because the relative importance of these effects is
difficult to quantify, the comparison of SFRs is most robust at
brighter $K$ magnitudes, and the agreement between the \Ha, X-ray, UV
and 24 \micron\ SFRs in this range is encouraging.  For the remaining
analysis, we adopt the corrected \Ha\ SFRs.

\subsection{Star Formation Rate Surface Density}
\label{sec:sfrd}
\begin{figure}[htbp]
\plotone{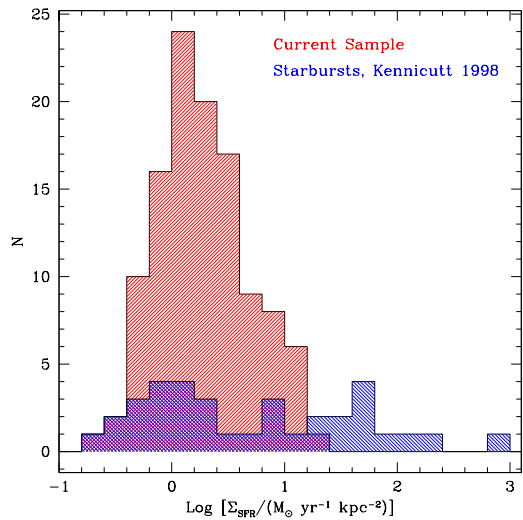}
\caption{A comparison of the star formation surface densities
  $\Sigma_{\rm SFR}$ of the current sample (large red histogram) and
  the starbursts of \citet[][shorter blue histogram]{k98schmidt}.  In
  this case we use a Salpeter IMF for consistency with the low
  redshift sample.  The inability to resolve star formation on small
  spatial scales at high redshift results in an absence of objects
  with the highest values of $\Sigma_{\rm SFR}$ in the \ztwo\ sample.}
\label{fig:sigsfr}
\end{figure}

Because we have measured the spatial extent of the \Ha\ emission (see
\citealt{ess+06}) as well as the star formation rate it implies, we
can also calculate the SFR surface density for the sample.  After
converting the SFRs to a Salpeter IMF by multiplying by 1.8 (for
comparison with local galaxies), we find a mean $\langle \Sigma_{\rm
  SFR} \rangle = 2.9$ \msunyr\ kpc$^{-2}$.  As shown in
Figure~\ref{fig:sigsfr}, the observed distribution is similar to the
sample of local starburst galaxies studied by \citet{k98schmidt}, with
the exception that the \ztwo\ sample does not contain objects with
$\Sigma_{\rm SFR} \gtrsim 20$ \msunyr\ kpc$^{-2}$; the upper cutoff of
our distribution is an order of magnitude lower than is seen locally.
The nearby galaxies with the highest values of $\Sigma_{\rm SFR}$ are
the ultra-luminous IR galaxies (ULIRGs), which have bolometric
luminosities $\gtrsim10^{12}$ L$_{\odot}$.  Recent
24 \micron\ observations from the Spitzer Space telescope have shown
that the most luminous \ztwo\ galaxies can be at least $\sim10$ times
more dust-obscured than would be inferred from their UV slopes
\citep{rsf+06,pmd+05}.  Thus it is possible that by using a UV-based
extinction correction we have underestimated the SFRs for the most
luminous galaxies in the sample, though by a smaller factor than we
would using UV-based SFRs because of the lower optical depth for \Ha.

However, the sample appears to include ULIRG-like objects.  From the
extinction-corrected \Ha\ SFRs we estimate that the bolometric
luminosities of the current sample range from $\sim10^{11}$ to
$\gtrsim 10^{12}$ L$_{\odot}$ (the bolometric luminosities inferred
from \Ha\ are plotted in Figure 10 of \citealt{ess+06}).  Most of the
local starbursts used by \citet{k98schmidt} are found in compact
circumnuclear disks, with sizes smaller than the galaxy in which they
are contained and smaller than the typical sizes we find for the
\ztwo\ galaxies.  It is not possible to resolve star formation on
scales smaller than a few kpc in the high redshift sample; starburst
activity that occurs in small, discrete regions rather than evenly
across the galaxy would lead to an overestimate of the size and an
underestimate of the surface density.

The large values of $\Sigma_{\rm SFR}$ imply high gas surface
densities and substantial gas masses.  This is discussed in detail by
\citet{ess+06}, in which we employ the correlation between
$\Sigma_{\rm SFR}$ and gas density to estimate the galaxies' gas
masses and gas fractions, finding a mean gas fraction of $\sim50$\%.
We also note that all of the objects have $\Sigma_{\rm SFR}>0.1$
\msunyr\ kpc$^{-2}$; starburst-driven superwinds are observed to be
ubiquitous in galaxies with SFR densities above this threshold
\citep{h02}.  The galaxies' outflow properties are discussed by
Steidel et al (2006, in preparation).

\section{Comparisons with Stellar Mass and Star Formation
  Timescales}
\label{sec:sftime}

\begin{figure*}[htbp]
\plotone{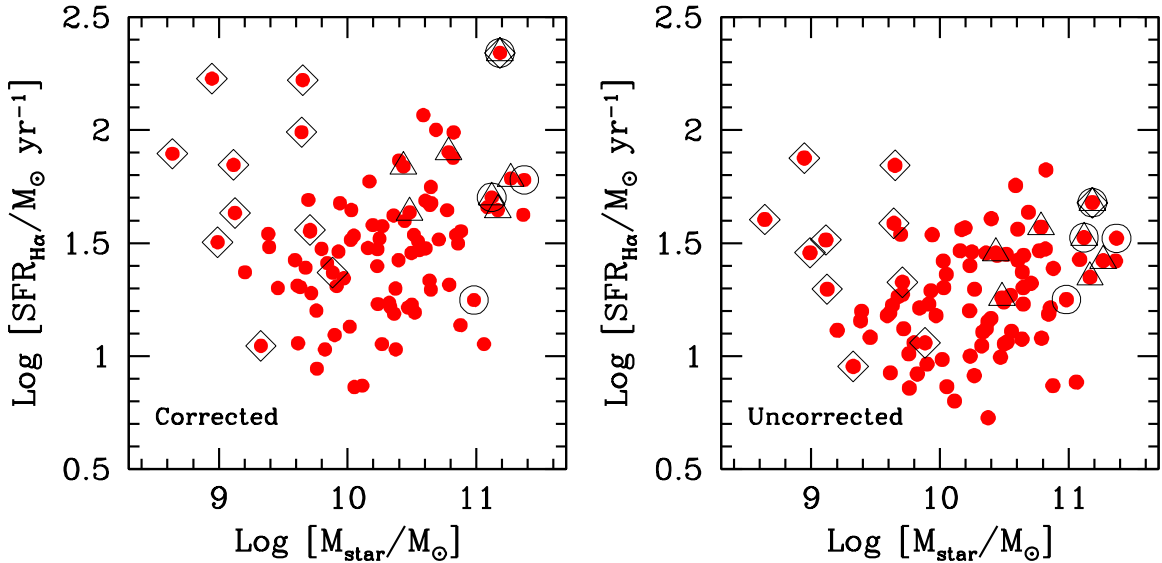}
\caption{Star formation rate from \Ha\ vs. stellar mass, with the SFR
  corrected for extinction at left and uncorrected at right.  In both
  cases SFR increases with increasing stellar mass, except for most of
  the galaxies with $M_{\rm dyn}/M_{\star}>10$ (marked with open
  diamonds; see \citealt{ess+06}).  The absence of low mass galaxies
  with low SFRs is probably a selection effect, as such objects are
  less likely to be detected both in our $K$-band images and in \Ha.
  Massive galaxies with little current star formation would also not
  be present in our survey.  Galaxies marked with triangles have
  $J-K>2.3$, and those marked with circles are AGN.}
\label{fig:sfrmass}
\end{figure*}

Given the suggestion of increasing SFR at brighter $K$ magnitudes
shown in Figure~\ref{fig:haxray} and found by \citet{res+05}, and the
correlation between stellar mass and $K$, we might expect a
correlation between SFR and stellar mass.  This is tested in
Figure~\ref{fig:sfrmass}, where in the left panel we show the
extinction-corrected $\rm SFR_{\Ha}$ plotted against
stellar mass.  There is a general trend in the sense that objects with
higher stellar masses have larger SFRs, but the data are only moderately
correlated with a significance of 2.1$\sigma$.  For the same set of
objects, $K$ magnitude and SFR are much more strongly correlated, with
4.3$\sigma$ significance; this is probably because the rest-frame
optical light is strongly affected by current star formation as well
as the formed stellar mass.

Some features of this plot can be explained by selection effects.  The
absence of objects with low stellar masses and low star formation
rates is probably due to the fact that we are less likely to detect
\Ha\ in galaxies that are faint in $K$.  A low mass galaxy would also
require a relatively high SFR to be detected in the observed $K$-band.
Massive, nearly passively evolving galaxies with low SFRs would also
not be selected by our survey.  This result can be usefully
compared with that of \citet{rsf+06}, who consider bolometric
luminosity as a function of stellar mass for optical and near-IR
selected galaxies (Figure 14).  They find that low mass galaxies span
a wide range in bolometric SFRs, from LIRG to ULIRG levels of
luminosity, and that the high mass and lower luminosity range of
parameter space contains galaxies selected with near-IR techniques;
thus, among galaxies of all types at \ztwo, the correlation between
stellar mass and SFR is relatively weak.

The points marked with open diamonds are objects in which the
dynamical mass $M_{\rm dyn}$, as determined by the \Ha\ line width and
the spatial extent of the \Ha\ emission, is more than 10 times larger
than the stellar mass $M_{\star}$.  Stellar and dynamical masses are
compared by \citet{ess+06}, who show that the galaxies with $M_{\rm
  dyn}/M_{\star}>10$ have young ages and high \Ha\ equivalent widths,
and are therefore likely to be young objects with large gas fractions.
This conclusion is further strengthened by estimates of their gas
masses, determined by making use of the correlation between star
formation rate surface density and gas density \citep{k98schmidt}; the
mean gas fraction implied for such $M_{\rm dyn} \gg M_{\star}$ objects
is $\sim90$\%.  These objects occupy a unique region in
Figure~\ref{fig:sfrmass}, with high SFRs and low stellar masses.

A possible concern is that the high SFRs of the young, low mass
objects are due to the extinction correction, if the degeneracy
between age and extinction has caused an overestimation of the
reddening.  In the right panel of Figure~\ref{fig:sfrmass} we plot the
uncorrected $\rm SFR_{\Ha}$ vs.\ stellar mass, two entirely
independently derived quantities; the plot is very similar to the
corrected version, with the $M_{\rm dyn}/M_{\star}>10$ objects still
among those with the highest SFRs in the sample.  A related concern is
that these young objects may not follow the same extinction law as the
rest of the sample; this is suggested by \citet{rsf+06}, who show that
(unlike most other UV-selected objects) galaxies with best fit ages
$<100$ Myr are offset from the local relation between the UV slope
$\beta$ and the ratio of far-IR to UV luminosity $L_{\rm
  FIR}/L_{1600}$.  If this is true, the extinction correction may be
overestimated for this set of objects.  However, the impact on our
results is negligible; we estimate that this could cause an
overestimate of the SFRs in young objects of a factor of $\sim1.2$,
significantly less than other sources of uncertainty.

\subsection{Star Formation Timescales}
\begin{figure}[htbp]
\plotone{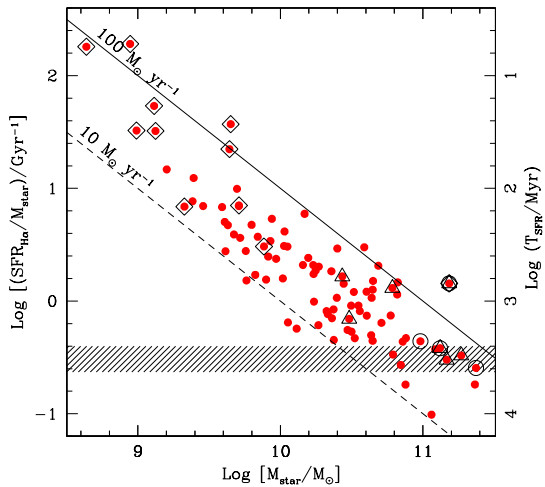}
\caption{The specific star formation rate SFR$_{\Ha}/M_{\star}$
  vs.\ stellar mass.  The solid and dashed diagonal lines show star
  formation rates of 100 and 10 \msunyr\ respectively.  The right axis
  shows the star formation rate timescale $T_{\rm SFR} =
  M_{\star}/{\rm SFR_{\Ha}}$, the inverse of the specific SFR.  On
  this scale, the shaded band represents the age of the universe for
  the redshift range of the galaxies in the sample.  The most massive
  galaxies have $T_{\rm SFR}\gtrsim t_{\rm universe}$, indicating that
  they may require declining star formation histories. Symbols are as
  in Figure~\ref{fig:sfrmass}.}
\label{fig:ssfr}
\end{figure}

A commonly used measure of the importance of the current episode of
star formation to the buildup of stellar mass in a galaxy is the
specific star formation rate, the star formation rate per unit stellar
mass.  Massive galaxies have lower specific SFRs, and at a given
stellar mass the specific SFR is observed to decline with redshift
(e.g.\ \citealt{pmd+05,rsf+06}).  We plot the specific SFR
against stellar mass in Figure~\ref{fig:ssfr}.  This plot can be
usefully compared with Figure 15 of \citet{rsf+06}, who plot specific
SFR as a function of stellar mass for both UV and near-IR selected
galaxies in the GOODS-N field.  The higher fraction of massive
galaxies in the NIRSPEC sample considered here shows that the
UV-selected sample contains objects with stellar masses and specific
SFRs comparable to the most massive near-IR selected objects with the
lowest specific SFRs in the GOODS-N field (neglecting those which are
not detected at 24 \micron).  Both figures show that at
\ztwo\ (as in the local universe), galaxies with low stellar masses are
assembling a much higher fraction of their stellar mass than more
massive objects.

The inverse of the specific SFR provides a star formation timescale,
$T_{\rm SFR}=M_{\star}/\rm SFR$; this is the time required for the
galaxy to form all its stellar mass at the current SFR.  By comparison
with the age of the universe at the redshift of the galaxy and with
the inferred age from the SED fits, we may obtain some constraints on
the star formation histories.  The right axis of Figure~\ref{fig:ssfr}
shows $T_{\rm SFR}$, and on this scale the shaded horizontal band
represents the age of the universe for the range of redshifts in the
sample.  If $T_{\rm SFR}$ is greater than the age of the universe at
the redshift of the galaxy, then the galaxy cannot have formed all its
stars at the current rate, and must have had a higher SFR in the past.
Only objects with $M_{\star} \gtrsim 6 \times 10^{10}$ \msun\ have
$T_{\rm SFR}$ approximately equal to the age of the universe.  This
upper limit on the time available for star formation suggests that
while most objects do not require declining star formation histories,
a CSF model may not be a reasonable fit for the most massive galaxies.
These appear from Figure~\ref{fig:ssfr} to require higher past SFRs,
although the uncertainties in the SFR and stellar masses are large
enough that this conclusion is not robust.  Similar results are found
from the SED modeling, as noted in \S\ref{sec:smass}; the issue is
discussed in more detail by \citet{sse+05}, who find that constant
star formation models do not provide an adequate fit to the SEDs of
five of the six galaxies in their sample with $M_{\star}>10^{11}$
\msun.  These galaxies, and the most massive objects in the current
sample, are best described by exponentially declining models with
$\tau=1$--2 Gyr.  With $\tau/t\gtrsim1$, such a model would be
indistinguishable from constant star formation for the younger
galaxies in the sample, and may in fact be preferred for these objects
because of the exponential SFR implied by a Schmidt law-like
dependence of the star formation rate on the gas mass (see the
discussion by \citealt{rsf+06}).

\begin{figure}[htbp]
\plotone{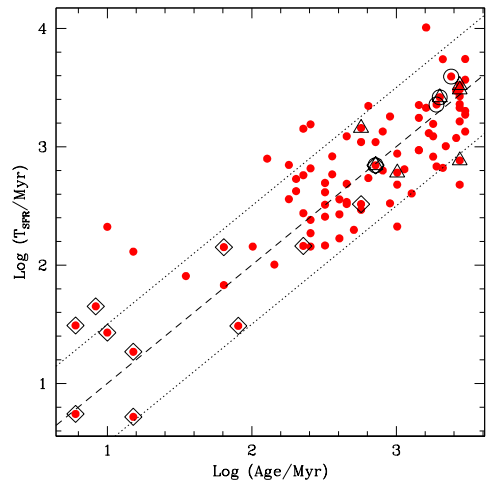}
\caption{The star formation timescale $T_{\rm SFR}$ vs.\ age.  This
  plot provides a check of the consistency of the \Ha\ SFRs and the
  primarily constant star formation models we use to fit the SEDs; for
  CSF models, $T_{\rm SFR}$ should be approximately equal to the age.
  The dashed line shows equal times, and the dotted lines on either
  side show the typical uncertainty in age.  Symbols are as in
  Figure~\ref{fig:sfrmass}.}
\label{fig:tsfr_age}
\end{figure}

We can obtain additional constraints on the star formation histories
by comparing $T_{\rm SFR}$ with the ages we obtain from the SED
fitting.  This test is implicit in the comparison of SFRs from
\Ha\ and the SED fitting shown in Figure~\ref{fig:sfrcomp}; it is
essentially a consistency check for our SED fits and \Ha\ SFRs, since
most of the ages represent constant star formation models.  If the
current SFR is an adequate representation of the past average, then
$T_{\rm SFR}$ should be approximately equal to the age. We plot
$T_{\rm SFR}$ vs.\ age in Figure~\ref{fig:tsfr_age}.  The dashed line
represents equal timescales; if objects fall significantly above this
line, they cannot have formed all of their stars at their current rate
over their inferred lifetime and must have had a past burst, while
objects significantly below the line would have a current SFR higher
than the past average.  The dotted lines show the average uncertainty
in the age from our Monte Carlo simulations of the SED fits, which
include uncertainties due to the star formation history. Most of the
objects fall between or near the dotted lines, suggesting that
constant star formation over the age determined by the SED fit
adequately describes the star formation histories of most of the
galaxies in our sample, though the scatter is certainly large enough
to allow for some declining star formation histories, as may be
required for the most massive galaxies.  It should also be noted that
the tendency of a few of the youngest galaxies to fall above the
dashed line is probably due to an underestimate of their ages, which
cannot realistically be less than their dynamical times; for this set
of objects, the average $t_{\rm dyn} \simeq 2r/\sigma = 80$ Myr (as
compared to $\sim130$ Myr for the entire sample).

\subsection{\Ha\ Equivalent Widths}
\begin{figure}[htbp]
\plotone{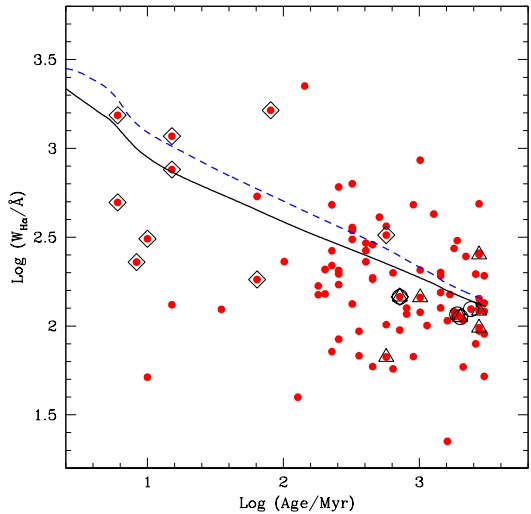}
\caption{A comparison of \Ha\ equivalent width and age from the SED
  modeling.  The lines show the predicted $W_{\Ha}$ as a function of
  age for constant star formation, from Starburst99 models with solar
  (solid black line) and 0.4 solar (dashed blue line) metallicity.
  The large scatter of the data with respect to the models is probably
  caused by variations in the SFR as well as observational
  uncertainties.  Symbols are as in Figure~\ref{fig:sfrmass}.}
\label{fig:ea_sb99}
\end{figure}

The \Ha\ equivalent width $W_{\Ha}$ provides an additional tool to
investigate the star formation history.  As the ratio of the
\Ha\ luminosity to the underlying stellar continuum, $W_{\Ha}$ is a
measure of the ratio of the current to past average star formation.
We determine $W_{\Ha}$ by taking the ratio of the \Ha\ flux and the
$K$-band continuum flux, after subtracting the contribution of \Ha\ to
the $K$-band magnitude.  In calculating the equivalent widths we have
applied the factor of two aperture correction to the \Ha\ fluxes
discussed above and in \S\ref{sec:irspec} (except in the cases of
Q1623-BX455 and Q1623-BX502, for which twice the \Ha\ flux slightly
exceeds the $K$-band magnitude), but we have not applied an extinction
correction; this is equivalent to the assumption that the nebular
emission lines and the stellar continuum suffer the same attenuation.
$W_{\Ha}$ is plotted against the best-fit age from the SED fits in
Figure~\ref{fig:ea_sb99}.  For constant star formation, $W_{\Ha}$
should decrease with age, as the stellar continuum increases while the
\Ha\ flux remains the same.  There is considerable scatter in the
$W_{\Ha}$--age comparison, but the probability that the data are
uncorrelated is $P=0.001$, for a significance of 3.3$\sigma$.

For simple star formation histories, the evolution of
$W_{\Ha}$ with galaxy age can be predicted with models of stellar evolution
and population synthesis.  The solid black line in
Figure~\ref{fig:ea_sb99} is the
theoretically predicted dependence of $W_{\Ha}$ on age, from a
Starburst99 \citep{sb99} model with constant star formation, solar
metallicity, and a \citet{k01} IMF, which gives very similar results
to the Chabrier IMF we employ; the dashed blue line is the same, but
for $Z=0.4Z_{\odot}$ (as discussed above, metal-rich galaxies are
observed to produce less \Ha\ luminosity for a given SFR than galaxies
of lower metallicity).  There is general agreement between the models and
the data, but with a large amount of scatter.  The equivalent width is
a comparison of two quantities with very different timescales; the
light from the stellar continuum generally increases over time, while
the \Ha\ flux may vary stochastically on a much shorter timescale, in
response to mergers, feedback, or accretion events.  The scatter in
the data with respect to the models is $\sim0.5$ dex, which can be
accounted for by a factor of $\sim2$ change in the current star formation rate
with respect to the past average (because a change in the \Ha\ flux also
affects the inferred continuum flux through the subtraction of \Ha,
the equivalent width can change by a larger factor than the star
formation rate).  A factor of $\sim2$ is also the typical uncertainty
in the star formation rate of individual objects.  

The relative extinction of the nebular lines and stellar continuum probably also
affects the results here.  As mentioned above in the discussion of the
star formation rates, we have not used the \citet{cab+00} prescription
of $E(B-V)_{\rm stellar}=0.4 E(B-V)_{\rm neb}$ for the extinction
corrections because doing so results in a
significant overestimate of the SFR$_{\Ha}$ with respect to the SFRs
from the UV continuum and our models (if we have overestimated the
typical aperture correction, then there is room for additional nebular
line extinction).  Applying this additional
extinction correction results in a typical increase of a factor of
$\sim3$ in $W_{\Ha}$; as can be seen in Figure~\ref{fig:ea_sb99}, the
mean value of $W_{\Ha}$ is somewhat below the CSF predictions at a
given age, but not usually by a factor of three.  It is possible that
the HII regions do suffer some smaller amount of additional
extinction, however, and this may explain the larger number of objects
in our sample that fall below the predictions.  This should be more
significant for the older objects, as the stars in young galaxies have
not had as much time to migrate away from the dusty regions in which
they form.  It appears from Figure~\ref{fig:ea_sb99}, however, that it
is the youngest
objects that have systematically lower equivalent widths than predicted by
the models.  These are also the objects for which $W_{\Ha}$ is the
most uncertain, however.  The typical uncertainty in $W_{\Ha}$ is
$\sim40$\%, but this approaches $\sim100$\% for the galaxies in which
the $\Ha$ flux makes up  most of the $K$-band light; comparisons of
ages and equivalent widths should be regarded as highly uncertain in this regime.
We also note that the most anomalous point, in the lower left corner,
corresponds to Q1700-BX681, which is not well fit by any model SED and
therefore has a very uncertain age.  As mentioned above, we may have somewhat
underestimated the ages of the youngest objects in general, as the ages cannot
be significantly less than the dynamical timescale $t_{\rm dyn}\sim80$ Myr.

Nothing in these results contradicts the hypothesis that the current star
formation rate is generally representative of the past average for
most of the sample, although stochastic variations are likely.  We are
not able to strongly discriminate between star formation histories,
however; a shallowly declining star formation history would also
result in equivalent widths somewhat below the CSF predictions, and
this is likely to be an additional factor for some of the objects in
the sample, particularly the oldest and most massive.  Extrapolating
forward in time, the star formation rates of the galaxies in our
sample will certainly decline as they lose their gas to star formation
or winds; by $z\sim1$ their clustering properties will best match
those of the early-type galaxies in the DEEP2 survey \citep{asp+05},
suggesting that star formation will be largely completed within the
next $\sim3$ Gyr.

\section{Summary and Discussion}
\label{sec:conclude}

We have used the \Ha\ and UV luminosities of a sample of 114 galaxies
at \ztwo\ in order to estimate their star formation rates.  Using
stellar masses and ages determined through population synthesis
modeling, we have assessed the star formation properties as a function
of stellar mass and age.  Our main conclusions are as
follows\footnote{Note that we have used a \citet{c03} IMF, which
  results in SFRs and stellar masses 1.8 times lower that the often
  used Salpeter IMF.}:

1.  The sample has a mean star formation rate from
extinction-corrected \Ha\ luminosity $\langle \rm SFR_{\Ha} \rangle =
31$ \msunyr.  The average extinction-corrected UV SFR is $\langle \rm
SFR_{UV} \rangle = 29$ \msunyr.  SFRs range from $\sim7$ to
$\gtrsim200$ \msunyr, and the average \Ha\ SFRs are in excellent agreement
with those determined from X-ray, radio, and mid-IR data.  The good
agreement between the indicators implies that the UV luminosity is
attenuated by an typical factor of $\sim4.5$, while the
\Ha\ luminosity is attenuated by a factor of $\sim1.7$ on average.  UV
attenuation ranges from none to a factor of $\gtrsim100$, and
\Ha\ attenuation from none to a factor of $\sim5$.

2.  Star formation rate and $K$ magnitude show significant
(4.3$\sigma$) correlation, with the brightest, $K_s<20$ galaxies
having $\langle \rm SFR_{\Ha} \rangle \sim 60$ \msunyr.  The
correlation between SFR and $K$ magnitude is significantly stronger
than the correlation between SFR and stellar mass, probably because
the rest-frame optical light is strongly affected by current star
formation as well as the formed stellar mass.

3.  All galaxies in the sample have SFRs per unit area $\Sigma_{\rm
  SFR}$ in the range observed in local starbursts.  All are also above
the threshold $\Sigma_{\rm SFR} \geq 0.1$ \msun\ yr$^{-1}$ kpc$^{-2}$ , above
which galactic-scale outflows are observed to be ubiquitous in the
local universe.

4.  We compare the instantaneous SFRs and the past average SFRs as
inferred from the ages and stellar masses, finding that for most of
the sample, the current SFR appears to be an adequate representation
of the past average.  There is some evidence that the most massive
galaxies ($M_{\star}>10^{11}$ \msun) have had higher SFRs in the
past.  Both of these conditions can be met by an exponentially
declining star formation rate with $\tau=1$--2 Gyr.

It is worth emphasizing the good overall agreement between SFRs
determined from \Ha, the UV continuum, X-rays, and radio and mid-IR
observations.  All of these diagnostics indicate the same average SFR
for the sample, and the dispersion between the \Ha\ and UV SFRs
suggests a typical uncertainty of a factor of $\sim2$.  This result
has encouraging implications for the determination of SFRs and the SFR
density at high redshift, as it is far easier to obtain UV
luminosities for a large sample of galaxies than \Ha\ fluxes or deep
mid-IR data (and radio and X-ray observations give only the average
SFRs of the sample).  There has been a widespread perception that the
UV luminosity is an unreliable measure of the instantaneous star
formation rate, but these results indicate that, for large numbers of
high redshift star-forming galaxies, this is not the case.

Another way of stating this result is that the UV slope provides a
reasonably accurate indication of extinction in most high redshift
star-forming galaxies.  This is not a new result; \citet{rs04} found
that UV luminosities uncorrected for extinction underestimated the
bolometric SFRs as determined from X-rays by a factor of $\sim4.5$--5,
in very good agreement with the factor of 4.5 difference we find
between the median corrected and uncorrected UV SFRs.  Using
bolometric luminosities determined from 24\micron\ fluxes,
\citet{rsf+06} find that most star-forming galaxies at \ztwo\ follow
the local relation between the rest-frame UV slope and dust
obscuration.  There are important exceptions to this rule, however;
the relationship between UV slope and obscuration breaks down for the
most luminous galaxies with $L_{\rm bol} \gtrsim 10^{12}$ L$_{\odot}$,
and young galaxies with ages less than $\sim100$ Myr also fall away
from the relation.

We have also found that, for most of the galaxies in the sample, the
current star formation rate appears to be representative of the past
average.  These results can be usefully compared with those of studies
at somewhat lower redshifts; for example, \citet{jgc+05} use galaxies
from the Gemini Deep Deep Survey (GDDS) in the redshift range
$0.8<z<2$ to study the dependence of star formation rate on stellar
mass.  They find that star formation in massive galaxies
($M_{\star}\sim6-30 \times 10^{10}$ \msun) drops steeply after
\ztwo\ and reaches its low present day value at $z\sim1$, while the
SFR declines more slowly in less massive galaxies.  In agreement with
this conclusion, we find that all of the galaxies in the current
sample are still strongly forming stars, and that the most massive
objects are likely to have had higher star formation rates in the
past.  \citet{ess+06} and \citet{rsf+06} show that these massive
galaxies probably have low gas fractions and have thus nearly finished
assembling their stellar mass.  More generally, the clustering
properties of the \ztwo\ galaxies \citep{asp+05} indicate that they
will become early-type galaxies with little current star formation by
$z\sim1$.

\acknowledgements We thank Andrew Blain, Jonathan Bird, David Kaplan
and Shri Kulkarni for obtaining near-IR images of some of our targets,
and the staffs of the Keck and Palomar observatories for their
assistance with the observations. We also thank the anonymous referee
for a useful report.  CCS, DKE and NAR have been supported
by grant AST03-07263 from the US National Science Foundation and by
the David and Lucile Packard Foundation.  AES acknowledges support
from the Miller Institute for Basic Research in Science, and KLA from
the Carnegie Institution of Washington.  Finally, we wish to extend
special thank s to those of Hawaiian ancestry on whose sacred mountain
we are privileged to be guests.  Without their generous hospitality,
most of the observations presented herein would not have been
possible.


\clearpage
\LongTables
\begin{landscape}
\begin{deluxetable}{l l l l c c c c c c c c c}
\tablewidth{0pt}
\tabletypesize{\footnotesize}
\tablecaption{Star Formation Rates\label{tab:sfr}}
\tablehead{
\colhead{} &
\colhead{} &
\colhead{} &
\colhead{} &
\colhead{} &
\colhead{Uncorrected} &
\colhead{Corrected} &
\colhead{} &
\colhead{} &
\colhead{Uncorrected} &
\colhead{Corrected} &
\colhead{} &
\colhead{} \\
\colhead{Object} & 
\colhead{$z_{\Ha}$} &
\colhead{$E(B-V)$\tablenotemark{a}} & 
\colhead{$F_{\Ha}$\tablenotemark{b}} & 
\colhead{$L_{\Ha}$\tablenotemark{c}} & 
\colhead{SFR$_{\Ha}$\tablenotemark{d}} & 
\colhead{SFR$_{\Ha}$\tablenotemark{e}} & 
\colhead{$m_{1500}$\tablenotemark{f}} &
\colhead{${\rm Log} (L_{1500})$\tablenotemark{g}} &
\colhead{SFR$_{\rm UV}$\tablenotemark{h}} &
\colhead{SFR$_{\rm UV}$\tablenotemark{i}} &
\colhead{SFR$_{\rm fit}$\tablenotemark{j}} &
\colhead{$W_{\Ha}$\tablenotemark{k}}
}
\startdata
CDFb-BN88 & 2.2615 & 0.149 & 2.6 & 1.0 & 9 & 14 & 23.43 & 29.27 & 14 & 60 & ... & ...\\
HDF-BX1055 & 2.4899 & 0.103 & 2.6 & 1.3 & 11 & 15 & 24.33 & 28.98 & 8 & 21 & 4 & 116\\
HDF-BX1084 & 2.4403 & 0.120 & 7.3 & 3.4 & 30 & 44 & 23.5 & 29.30 & 16 & 51 & ... & ...\\
HDF-BX1085 & 2.2407 & 0.171 & 1.1 & 0.4 & 4 & 6 & 24.83 & 28.70 & 4 & 20 & ... & ...\\
HDF-BX1086 & 2.4435 & 0.196 & 1.8 & 0.8 & 7 & 14 & 25.05 & 28.68 & 4 & 26 & ... & ...\\
HDF-BX1277 & 2.2713 & 0.095 & 5.3 & 2.1 & 18 & 25 & 24.01 & 29.04 & 8 & 21 & ... & ...\\
HDF-BX1303 & 2.3003 & 0.100 & 2.6 & 1.0 & 9 & 12 & 24.83 & 28.72 & 4 & 11 & 15 & 308\\
HDF-BX1311 & 2.4843 & 0.105 & 8.0 & 3.9 & 34 & 48 & 23.5 & 29.31 & 16 & 46 & 7 & 101\\
HDF-BX1322 & 2.4443 & 0.085 & 2.0 & 0.9 & 8 & 11 & 24.03 & 29.09 & 10 & 22 & 34 & 197\\
HDF-BX1332 & 2.2136 & 0.290 & 4.4 & 1.6 & 14 & 35 & 23.96 & 29.04 & 8 & 135 & 19 & 68\\
HDF-BX1368 & 2.4407 & 0.160 & 8.8 & 4.1 & 36 & 59 & 24.09 & 29.06 & 9 & 44 & 159 & 132\\
HDF-BX1376 & 2.4294 & 0.070 & 2.2 & 1.0 & 9 & 11 & 24.49 & 28.90 & 6 & 12 & 37 & 266\\
HDF-BX1388 & 2.0317 & 0.265 & 5.8 & 1.8 & 15 & 34 & 24.82 & 28.63 & 3 & 38 & 9 & 265\\
HDF-BX1397 & 2.1328 & 0.150 & 5.3 & 1.8 & 16 & 25 & 24.26 & 28.89 & 6 & 25 & 23 & 90\\
HDF-BX1409 & 2.2452 & 0.290 & 8.5 & 3.2 & 29 & 69 & 25.15 & 28.57 & 3 & 47 & 17 & 207\\
HDF-BX1439 & 2.1865 & 0.175 & 8.8 & 3.2 & 28 & 48 & 24.16 & 28.95 & 7 & 36 & 27 & 145\\
HDF-BX1479 & 2.3745 & 0.110 & 2.5 & 1.1 & 10 & 14 & 24.55 & 28.86 & 6 & 16 & 21 & 107\\
HDF-BX1564 & 2.2225 & 0.065 & 8.6 & 3.2 & 28 & 34 & 23.55 & 29.21 & 13 & 23 & 13 & 126\\
HDF-BX1567 & 2.2256 & 0.050 & 4.0 & 1.5 & 13 & 15 & 23.68 & 29.15 & 11 & 18 & 9 & 93\\
HDF-BX305 & 2.4839 & 0.285 & 4.2 & 2.1 & 18 & 43 & 25.07 & 28.68 & 4 & 65 & 5 & 72\\
HDF-BMZ1156\tablenotemark{l} & 2.2151 & 0.000 & 5.4 & 2.0 & 18 & 18 & 24.61 & 28.78 & 5 & 5 & 53 & 67\\
Q0201-B13 & 2.1663 & 0.003 & 2.4 & 0.8 & 7 & 8 & 23.36 & 29.26 & 14 & 15 & ... & ...\\
Q1307-BM1163 & 1.4105 & 0.178 & 28.7 & 3.5 & 31 & 53 & 22.21 & 29.38 & 19 & 99 & ... & ...\\
Q1623-BX151\tablenotemark{l} & 2.4393 & 0.059 & 3.5 & 1.6 & 14 & 17 & 24.74 & 28.80 & 5 & 9 & ... & ...\\
Q1623-BX214 & 2.4700 & 0.182 & 5.3 & 2.6 & 23 & 39 & 24.45 & 28.92 & 7 & 40 & ... & ...\\
Q1623-BX215 & 2.1814 & 0.134 & 4.8 & 1.7 & 15 & 22 & 24.71 & 28.73 & 4 & 15 & ... & ...\\
Q1623-BX252 & 2.3367 & 0.031 & 1.2 & 0.5 & 4 & 5 & 25.13 & 28.61 & 3 & 4 & ... & ...\\
Q1623-BX274 & 2.4100 & 0.119 & 9.5 & 4.3 & 38 & 54 & 23.48 & 29.29 & 15 & 50 & ... & ...\\
Q1623-BX344 & 2.4224 & 0.189 & 17.1 & 7.9 & 69 & 123 & 24.81 & 28.77 & 5 & 30 & ... & ...\\
Q1623-BX366 & 2.4204 & 0.200 & 7.9 & 3.6 & 32 & 58 & 24.25 & 28.99 & 8 & 55 & ... & ...\\
Q1623-BX376 & 2.4085 & 0.175 & 5.3 & 2.4 & 21 & 36 & 23.55 & 29.27 & 14 & 81 & 80 & 183\\
Q1623-BX428 & 2.0538 & 0.000 & 2.7 & 0.8 & 7 & 7 & 24.08 & 28.93 & 7 & 7 & 1 & 84\\
Q1623-BX429 & 2.0160 & 0.120 & 5.1 & 1.5 & 13 & 19 & 23.75 & 29.05 & 9 & 26 & 23 & 219\\
Q1623-BX432 & 2.1817 & 0.060 & 5.4 & 1.9 & 17 & 20 & 24.68 & 28.74 & 4 & 8 & 6 & 427\\
Q1623-BX447 & 2.1481 & 0.050 & 5.6 & 1.9 & 17 & 20 & 24.65 & 28.74 & 4 & 7 & 5 & 154\\
Q1623-BX449 & 2.4185 & 0.110 & 3.5 & 1.6 & 14 & 20 & 25.06 & 28.67 & 4 & 11 & 9 & 196\\
Q1623-BX452 & 2.0595 & 0.195 & 4.4 & 1.4 & 12 & 22 & 24.93 & 28.60 & 3 & 19 & 14 & 121\\
Q1623-BX453 & 2.1816 & 0.275 & 13.8 & 4.9 & 43 & 100 & 23.86 & 29.07 & 9 & 123 & 107 & 187\\
Q1623-BX455 & 2.4074 & 0.265 & 18.8 & 8.6 & 75 & 169 & 25.15 & 28.63 & 3 & 45 & 58 & 1172\tablenotemark{m}\\
Q1623-BX458 & 2.4194 & 0.165 & 4.3 & 2.0 & 17 & 29 & 23.69 & 29.21 & 13 & 65 & 55 & 102\\
Q1623-BX472 & 2.1142 & 0.130 & 3.9 & 1.3 & 11 & 17 & 24.74 & 28.69 & 4 & 13 & 11 & 135\\
Q1623-BX502 & 2.1558 & 0.220 & 13.2 & 4.6 & 40 & 79 & 24.57 & 28.77 & 5 & 37 & 72 & 1536\tablenotemark{m}\\
Q1623-BX511 & 2.2421 & 0.235 & 3.4 & 1.3 & 11 & 23 & 25.79 & 28.32 & 2 & 15 & 13 & 325\\
Q1623-BX513 & 2.2473 & 0.145 & 3.3 & 1.3 & 11 & 17 & 23.51 & 29.23 & 13 & 53 & 46 & 59\\
Q1623-BX516 & 2.4236 & 0.145 & 5.2 & 2.4 & 21 & 33 & 24.24 & 28.99 & 8 & 32 & 28 & 112\\
Q1623-BX522 & 2.4757 & 0.180 & 2.8 & 1.4 & 12 & 21 & 24.81 & 28.78 & 5 & 28 & 24 & 79\\
Q1623-BX528 & 2.2682 & 0.175 & 7.7 & 3.0 & 27 & 46 & 23.81 & 29.12 & 10 & 55 & 44 & 94\\
Q1623-BX543 & 2.5211 & 0.305 & 8.6 & 4.4 & 39 & 98 & 23.55 & 29.30 & 16 & 336 & 528 & 229\\
Q1623-BX586 & 2.1045 & 0.195 & 5.1 & 1.7 & 15 & 27 & 24.9 & 28.62 & 3 & 20 & 17 & 192\\
Q1623-BX599 & 2.3304 & 0.125 & 18.1 & 7.6 & 67 & 98 & 23.66 & 29.20 & 12 & 42 & 35 & 303\\
Q1623-BX663\tablenotemark{l} & 2.4333 & 0.135 & 8.2 & 3.8 & 33 & 50 & 24.38 & 28.94 & 7 & 26 & 21 & 112\\
Q1623-MD107 & 2.5373 & 0.060 & 3.7 & 1.9 & 17 & 20 & 25.47 & 28.54 & 3 & 5 & 4 & 858\\
Q1623-MD66 & 2.1075 & 0.235 & 19.7 & 6.5 & 57 & 116 & 24.32 & 28.86 & 6 & 50 & 43 & 482\\
Q1700-BX490 & 2.39597 & 0.285 & 17.7 & 8.0 & 70 & 166 & 23.24 & 29.39 & 19 & 313 & 448 & 310\\
Q1700-BX505 & 2.3089 & 0.270 & 3.6 & 1.5 & 13 & 29 & 25.62 & 28.41 & 2 & 27 & 20 & 121\\
Q1700-BX523 & 2.4756 & 0.260 & 4.7 & 2.3 & 20 & 44 & 24.97 & 28.72 & 4 & 55 & 42 & 171\\
Q1700-BX530 & 1.94294 & 0.045 & 12.2 & 3.3 & 29 & 33 & 23.26 & 29.22 & 13 & 19 & 6 & 208\\
Q1700-BX536 & 1.9780 & 0.115 & 11.3 & 3.2 & 28 & 40 & 23.21 & 29.25 & 14 & 40 & 15 & 150\\
Q1700-BX561 & 2.4332 & 0.130 & 1.9 & 0.9 & 8 & 11 & 24.84 & 28.76 & 4 & 16 & 10 & 22\\
Q1700-BX581 & 2.4022 & 0.215 & 4.0 & 1.8 & 16 & 30 & 24.15 & 29.02 & 8 & 69 & 70 & 124\\
Q1700-BX681 & 1.73959 & 0.315 & 6.3 & 1.3 & 11 & 30 & 22.23 & 29.54 & 27 & 427 & 628 & 52\\
Q1700-BX691 & 2.1895 & 0.125 & 7.7 & 2.8 & 24 & 36 & 25.55 & 28.39 & 2 & 6 & 5 & 257\\
Q1700-BX717 & 2.4353 & 0.090 & 3.8 & 1.8 & 16 & 20 & 24.98 & 28.70 & 4 & 10 & 8 & 410\\
Q1700-BX759 & 2.4213 & 0.230 & 1.3 & 0.6 & 5 & 11 & 24.79 & 28.77 & 5 & 45 & 37 & 57\\
Q1700-BX794 & 2.2473 & 0.130 & 6.8 & 2.6 & 23 & 34 & 23.95 & 29.05 & 9 & 31 & 25 & 183\\
Q1700-BX917 & 2.3069 & 0.040 & 7.4 & 3.0 & 27 & 30 & 24.71 & 28.77 & 5 & 7 & 4 & 117\\
Q1700-MD69 & 2.2883 & 0.275 & 7.5 & 3.0 & 26 & 61 & 25.22 & 28.56 & 3 & 40 & 31 & 122\\
Q1700-MD94\tablenotemark{l} & 2.33624 & 0.500 & 12.9 & 5.4 & 48 & 219 & 25.66 & 28.40 & 2 & 253 & 213 & 146\\
Q1700-MD103 & 2.3148 & 0.305 & 8.2 & 3.4 & 30 & 76 & 24.69 & 28.78 & 5 & 90 & 65 & 120\\
Q1700-MD109 & 2.2942 & 0.175 & 2.8 & 1.1 & 10 & 17 & 25.72 & 28.36 & 2 & 10 & 8 & 246\\
Q1700-MD154\tablenotemark{l} & 2.62911 & 0.335 & 4.1 & 2.3 & 20 & 56 & 23.96 & 29.17 & 12 & 359 & 347 & 40\\
Q1700-MD174 & 2.3423 & 0.195 & 8.9 & 3.8 & 33 & 60 & 24.88 & 28.71 & 4 & 27 & 24 & 125\\
Q2343-BM133 & 1.47744 & 0.115 & 28.7 & 3.9 & 35 & 49 & 22.78 & 29.19 & 12 & 36 & 35 & 2245\\
Q2343-BM181 & 1.4951 & 0.134 & 3.4 & 0.5 & 4 & 6 & 25.18 & 28.24 & 1 & 5 & ... & ...\\
Q2343-BX163 & 2.12132 & 0.050 & 2.2 & 0.7 & 6 & 7 & 24.06 & 28.97 & 7 & 12 & 9 & 127\\
Q2343-BX169 & 2.20939 & 0.125 & 4.7 & 1.7 & 15 & 22 & 23.3 & 29.30 & 16 & 51 & 46 & 152\\
Q2343-BX182 & 2.2879 & 0.100 & 2.4 & 1.0 & 8 & 11 & 23.88 & 29.10 & 10 & 26 & 23 & 168\\
Q2343-BX236 & 2.43475 & 0.085 & 3.1 & 1.4 & 13 & 16 & 24.42 & 28.93 & 7 & 15 & 13 & 150\\
Q2343-BX336 & 2.54387 & 0.210 & 4.3 & 2.2 & 20 & 38 & 24.31 & 29.00 & 8 & 66 & 58 & 133\\
Q2343-BX341 & 2.5749 & 0.210 & 4.0 & 2.1 & 19 & 36 & 24.59 & 28.90 & 6 & 52 & 50 & 231\\
Q2343-BX378 & 2.04407 & 0.165 & 4.5 & 1.4 & 12 & 20 & 25.06 & 28.54 & 3 & 12 & 11 & 606\\
Q2343-BX389 & 2.17156 & 0.250 & 12.0 & 4.2 & 37 & 80 & 25.13 & 28.56 & 3 & 30 & 22 & 253\\
Q2343-BX390 & 2.2313 & 0.150 & 4.9 & 1.9 & 16 & 26 & 24.6 & 28.79 & 5 & 20 & 17 & 293\\
Q2343-BX391 & 2.17403 & 0.195 & 4.2 & 1.5 & 13 & 24 & 24.51 & 28.80 & 5 & 31 & 25 & 537\\
Q2343-BX418 & 2.3052 & 0.035 & 8.0 & 3.3 & 29 & 32 & 23.94 & 29.08 & 9 & 13 & 12 & 1639\\
Q2343-BX429 & 2.1751 & 0.185 & 4.8 & 1.7 & 15 & 27 & 25.42 & 28.44 & 2 & 12 & 12 & 632\\
Q2343-BX435 & 2.1119 & 0.225 & 8.1 & 2.7 & 24 & 47 & 24.61 & 28.74 & 4 & 35 & 30 & 200\\
Q2343-BX436 & 2.3277 & 0.070 & 7.2 & 3.0 & 26 & 33 & 23.19 & 29.38 & 19 & 37 & 33 & 345\\
Q2343-BX442 & 2.1760 & 0.225 & 7.2 & 2.5 & 22 & 44 & 24.48 & 28.82 & 5 & 43 & 25 & 98\\
Q2343-BX461 & 2.5662 & 0.250 & 7.0 & 3.7 & 33 & 70 & 24.84 & 28.80 & 5 & 62 & 86 & 760\\
Q2343-BX474 & 2.2257 & 0.215 & 5.0 & 1.9 & 16 & 32 & 24.73 & 28.73 & 4 & 33 & 26 & 133\\
Q2343-BX480 & 2.2313 & 0.165 & 3.0 & 1.1 & 10 & 16 & 24.06 & 29.00 & 8 & 38 & 33 & 67\\
Q2343-BX493 & 2.33964 & 0.255 & 5.3 & 2.2 & 20 & 43 & 23.91 & 29.10 & 10 & 118 & 220 & 497\\
Q2343-BX513 & 2.10919 & 0.135 & 10.1 & 3.3 & 29 & 44 & 24.13 & 28.93 & 7 & 24 & 20 & 192\\
Q2343-BX529 & 2.1129 & 0.145 & 3.5 & 1.2 & 10 & 16 & 24.62 & 28.74 & 4 & 17 & 14 & 230\\
Q2343-BX537 & 2.3396 & 0.130 & 5.2 & 2.2 & 19 & 29 & 24.67 & 28.80 & 5 & 17 & 15 & 365\\
Q2343-BX587 & 2.2430 & 0.180 & 5.5 & 2.1 & 19 & 32 & 23.79 & 29.12 & 10 & 57 & 49 & 95\\
Q2343-BX599 & 2.01156 & 0.100 & 4.5 & 1.3 & 12 & 16 & 23.6 & 29.11 & 10 & 25 & 21 & 107\\
Q2343-BX601 & 2.3769 & 0.125 & 7.4 & 3.3 & 29 & 42 & 23.7 & 29.20 & 12 & 42 & 36 & 199\\
Q2343-BX610 & 2.2094 & 0.155 & 8.1 & 3.0 & 26 & 42 & 23.92 & 29.05 & 9 & 38 & 32 & 59\\
Q2343-BX660 & 2.1735 & 0.010 & 9.4 & 3.3 & 29 & 30 & 24.27 & 28.90 & 6 & 7 & 5 & 488\\
Q2343-MD59 & 2.01159 & 0.200 & 2.9 & 0.8 & 7 & 14 & 24.99 & 28.55 & 3 & 18 & 11 & 52\\
Q2343-MD62 & 2.17524 & 0.150 & 2.3 & 0.8 & 7 & 11 & 25.5 & 28.41 & 2 & 8 & 7 & 143\\
Q2343-MD80 & 2.0138 & 0.020 & 3.2 & 0.9 & 8 & 9 & 24.81 & 28.63 & 3 & 4 & 1 & 206\\
Q2346-BX120 & 2.2664 & 0.005 & 5.3 & 2.1 & 18 & 19 & 25.1 & 28.60 & 3 & 3 & ... & ...\\
Q2346-BX220 & 1.9677 & 0.055 & 10.3 & 2.9 & 25 & 30 & 23.86 & 28.99 & 8 & 13 & 4 & 482\\
Q2346-BX244 & 1.6465 & 0.300 & 5.4 & 1.0 & 9 & 21 & 23.49 & 29.00 & 8 & 149 & ... & ...\\
Q2346-BX404 & 2.0282 & 0.095 & 13.9 & 4.2 & 36 & 49 & 23.57 & 29.13 & 10 & 25 & 22 & 273\\
Q2346-BX405 & 2.0300 & 0.010 & 14.0 & 4.2 & 37 & 38 & 23.44 & 29.18 & 12 & 13 & 7 & 358\\
Q2346-BX416 & 2.2404 & 0.195 & 12.1 & 4.6 & 41 & 73 & 23.89 & 29.08 & 9 & 60 & 55 & 287\\
Q2346-BX482 & 2.2569 & 0.112 & 11.2 & 4.4 & 38 & 54 & 23.54 & 29.22 & 13 & 38 & ... & ...\\
SSA22a-MD41 & 2.1713 & 0.096 & 7.9 & 2.8 & 25 & 33 & 23.5 & 29.21 & 13 & 31 & ... & ...\\
West-BM115 & 1.6065 & 0.225 & 5.9 & 1.0 & 9 & 17 & 24.05 & 28.75 & 4 & 40 & ... & ...\\
West-BX600 & 2.1607 & 0.047 & 6.3 & 2.2 & 19 & 22 & 24.04 & 28.99 & 8 & 12 & ... & ...
\enddata

\tablenotetext{a}{$E(B-V)$ inferred from SED fitting when $K$-band
  photometry is present (indicated by a value in column 11, the SFR
  from SED fitting), and
  calculated from the $G-{\cal R}$ color assuming an SED with constant
  star formation and an age of 570 Myr otherwise.}
\tablenotetext{b}{Observed flux of \Ha\ emission line, in units of $10^{-17}$ erg
  s$^{-1}$ cm$^{-2}$.}
\tablenotetext{c}{Observed \Ha\ luminosity, in units of $10^{42}$ erg s$^{-1}$.}
\tablenotetext{d}{SFR derived from \Ha\ flux in \msunyr, uncorrected
  for extinction and applying a factor of two aperture correction.}
\tablenotetext{e}{SFR derived from \Ha\ flux after correcting for
  extinction and slit losses, in \msunyr.}
\tablenotetext{f}{Observed magnitude at $\sim1500$ \AA; $G$-band for
  most objects, $U_n$ for those with $z\sim1.5$.}
\tablenotetext{g}{Observed rest frame UV luminosity, log (erg s$^{-1}$
  cm$^{-2}$ Hz$^{-1}$).}
\tablenotetext{h}{SFR derived from uncorrected UV magnitude, in \msunyr.}
\tablenotetext{i}{SFR derived from extinction-corrected UV magnitude, in \msunyr.}
\tablenotetext{j}{SFR derived from SED fitting, in \msunyr.}
\tablenotetext{k}{\Ha\ equivalent width in \AA, incorporating a factor
  of two aperture correction except where noted.}
\tablenotetext{l}{AGN, as determined from rest-frame UV or optical spectra.}
\tablenotetext{m}{Aperture correction not applied for equivalent width calculation.}
\end{deluxetable}

\clearpage
\end{landscape}

\end{document}